\documentclass[twocolumn,superscriptaddress,amssymb,amsmath,aps,prl]{revtex4-2}

\usepackage{color}
\usepackage{comment}
\usepackage{graphicx}
\usepackage{tensor}
\usepackage{float}

\newcommand{\rl}{r_{\text{iLR}}}

\newcommand{\Tf}{T^{\text{\scriptsize fluid}}}
\newcommand{\Tq}{T^{\text{\scriptsize quan}}}

\begin{document} 

\title{ Self-consistent solution to the semiclassical Einstein equations of a star}

\author{Pietro Paolo Melella}
\author{Ignacio A. Reyes}
\affiliation{Institute for Theoretical Physics, University of Amsterdam, PO Box 94485, 1090 GL Amsterdam, The Netherlands}

\begin{abstract}

We present the interior solution for a static, spherically symmetric perfect fluid star backreacted by QFT in four dimensions invoking no arbitrary parameters. It corresponds to a constant energy density star and is fully non-perturbative. The space of solutions includes ultra-compact configurations that have neither singularities nor light rings inside the star and can exist arbitrarily close to the Schwarzschild limit, showing that the classical paradigm of astrophysics does not hold once QFT in curved space is taken into account.

 \end{abstract}

\maketitle

\section{Introduction}

The current paradigm of Astrophysics is an entangled story between Quantum mechanics and General Relativity. It was essentially set by the work of Oppenheimer and Volkoff in the 1930's who, following Chandrasekhar's idea\,\cite{1931ApJ....74...81C}, used the quantum statistical mechanics of the free Fermi gas as the source for solving the Einstein's equations of a star\,\cite{PhysRev.55.374}. The essence of their result is that \textit{there exists no static and smooth solutions arbitrarily close to the Schwarzschild limit.} 

What this picture misses is to account for QFT coupled to the gravitational field. QFT exhibits physical effects that have no counterpart in relativistic quantum statistical mechanics. An essential instance of this, as we shall exploit here, are Anomalies. In this paper we present the first exact self-consistent solution of a star that takes into account the backreaction effects of QFT, and ask whether the conclusions of the ruling paradigm still hold.

We begin with a review of the classical interior Schwarzschild solution of the constant density star for two main reasons. First, it demonstrates how to find a solution using a symmetry instead of an equation of state. Second, it is a great solvable example to illustrate the current paradigm of astrophysics. Plus, when we consider the QFT backreaction it will again play an essential role.

Consider a static and spherically symmetric spacetime parametrized as:
\begin{align}\label{metric}
ds^2=-e^{2\nu(r)}dt^2+e^{2\lambda(r)}dr^2+r^2 d\Omega_2^2 \,,
\end{align}
coupled to matter described by a perfect fluid:
\begin{align}\label{Tmunuclass}
T^{\mu}{}_{\nu}{}{}^{\text{\scriptsize fluid}} = \text{diag}(-\rho,p,p,p)\,.
\end{align}

The paradigmatic example is the free Fermi gas at zero temperature\,\cite{1931ApJ....74...81C,PhysRev.55.374}. More generally, \eqref{Tmunuclass} together with its equation of state should emerge from a QCD calculation in flat space. This is of course a quantum calculation, but it is derived via QFT in \textit{Minkowski} space, where the expectation value in the vacuum state $|\Omega\rangle$ vanishes by definition: $\langle \Omega|T_{\mu\nu}{}^{\text{\scriptsize fluid}}|\Omega\rangle=0$. This does not hold in curved space. 

The equations of motion read:
\begin{align}\label{EEs0}
G_{\mu\nu} &= 8\pi G\, T_{\mu\nu}{}^{\text{\scriptsize fluid}}\,.
\end{align}

At this point we can count degrees of freedom. Due to the symmetries, \eqref{EEs0} reduce to three equations for the four unknowns ($\rho,p,\nu,\lambda$), meaning we are missing one equation. The usual approach consists of providing an equation of state relating $\rho$ and $p$. This is a reasonable strategy when one has more information about the microscopic constituents of the fluid, but it is not mandatory. A different approach is to impose some geometric condition, for example a symmetry. Then, if solutions are found, one can evaluate the density and pressure and \textit{deduce} the corresponding equation of state and consider its physical meaning. 

\subsection{Weyl flat solutions}

The condition we will use in this paper is to restrict to conformally flat metrics for the interior, which is equivalent to the vanishing of the Weyl tensor. We will mostly deal with the interior semiclassical solution and cannot determine the exterior self-consistent solution, although we uncover some of its properties.

Again due to the simplicity of the metric, it is easy to see that all non-zero components of the Weyl tensor are proportional to a single function, $W_{\mu\nu\rho\sigma}\propto W$, so the metric is conformally flat if
\begin{align}\label{noweyl}
W= r^2 \nu''+  r(\nu'-\lambda') (r \nu'-1) -e^{2\lambda}+1=0\,.
\end{align}

This provides the remaining fourth equation to close the system. As proved by Buchdahl\,\cite{10.1119/1.1986083}, the only static distribution of fluid with non-negative density and pressure which generates a conformally flat metric through the classical Einstein equations without cosmological constant is the Schwarzschild interior solution. 

A convenient system of two equations for two unknowns is obtained by subtracting the $rr$ and $\theta\theta$ components of \eqref{EEs0} which eliminates the pressure, and imposing \eqref{noweyl} yields 
\begin{align}\label{eq000}
1-e^{2\lambda}+r\lambda'=0\,.
\end{align}
This, together with \eqref{noweyl}, is a set of two ODEs for the two unknown metric functions $\nu(r)$ and $\lambda(r)$. Finally the appropriate boundary conditions for local smoothness at the origin are:
\begin{align}\label{regularity}
    \lambda(0)=\lambda'(0)=\nu'(0)=0\,.
\end{align}

The solution is textbook material and can be parametrized as:
\begin{align}\label{lambda}
e^{-2 \lambda}&= 1-\frac{2GM_0 r^2}{R_0^3}\\
e^{\nu} &=  \frac{3}{2}\sqrt{1-\frac{2GM_0}{R_0}} - \frac{1}{2} \sqrt{1-\frac{2GM_0r^2}{R_0^3}} \label{nu}  
\end{align}
together with
\begin{align}\label{rho0}
\rho_0&=\frac{M_0}{\frac{4}{3}\pi R_0^3}\\
p_0(r)&=\rho_0 \frac{ \sqrt{R_0^3-2GM_0R_0^2} - \sqrt{R_0^3-2GM_0r^2}  }{\sqrt{R_0^3-2GM_0r^2}-3\sqrt{R_0^3-2GM_0R_0^2}}\label{p0}\,.
\end{align}

In this classical solution $M_0,R_0$ are two positive constants which represent the mass and radius of the sphere. This comes from the boundary conditions at the surface of the star. In the simplest case, one glues this interior to an exterior Schwarzschild vacuum solution
\begin{align}
    ds^2 = -\left( 1-\frac{2GM}{r} \right) dt^2 + \left( 1- \frac{2GM}{r} \right)^{-1}dr^2+r^2d\Omega_2^2\nonumber
\end{align}
where the fluid density and pressure $\rho_0,p_0$ vanish. Then by general physical principles, one requires that the pressure (not the density) is continuous at the boundary so the radius of the star is defined by
\begin{align}\label{pR0=0}
    p(R_0)=0\,,
\end{align}
as can be readily verified in \eqref{p0}. These boundary conditions at the surface imply that $\lambda'$ is discontinuous while $\nu'$ is not. The Riemann tensor, containing no $\lambda''$, is thus discontinuous but finite there. As usual, defining the Misner mass via $e^{-2\lambda}:=1-\frac{2G m(r)}{r}$, the Einstein equations give $m(r)=4\pi \int_0^r dr\, r^2 \rho$, and the total mass is then $M_0:=m(R_0)$.

\subsection{The paradigm}

The constant density star solution described above is a great illustration of the ruling paradigm alluded to earlier, namely \textit{that there is a gap between the most compact stars in the universe, and black holes, and nothing can live in between}. In classical GR there are two main results that suggest that ultracompact objets cannot exist, arguing they are either singular (spacetime is not regular) or unstable (e.g. under gravitational perturbations). More precisely, for a star of positive mass there exist no solutions both regular and gravitationally stable that are arbitrarily close to the black hole limit. This picture relies only on the following theorems.

Let's consider first the issue of smoothness. Buchdahl proved\,\cite{Buchdahl:1959zz} that for a sphere of radius $R_0$ and mass $M_0$ made of perfect fluid, if $\rho>0$ and $\partial_r \rho\leq 0$ but assuming nothing about the equation of state, static spherically symmetric solutions to the Einstein equations \eqref{EEs0} are regular (free of singularities) only if they are above the Buchdahl bound
\begin{align}
    \frac{R_0}{GM_0}> \frac{9}{4}\,.
\end{align}

The constant density solution is a perfect example of this: the central pressure \eqref{p0} diverges as we approach the bound. 
Beyond the Buchdahl limit, \eqref{lambda}-\eqref{nu} are still exact static solutions to \eqref{metric}-\eqref{EEs0}, but they all have curvature singularities. Therefore, under these assumptions, solutions become singular much before we reach the Schwarzschild radius.  

The second set of results\,\cite{Keir:2014oka,Cunha:2017qtt, DiFilippo:2024ddg} argues for instability, which is linked to the existence of light rings -- trajectories where null rays form closed orbits in space. This can be understood by recasting the null geodesic radial equation in the form $e^{2(\nu+\lambda)}\dot r^2+V=E$, where $V$ is the effective potential. For compact enough solutions with $R_0<3GM_0$, two light rings -- extrema of $V$ -- appear. If the exterior is the Schwarzschild vacuum metric, the outer light ring always sits at $r=3GM_0$, whereas the inner one is inside the surface of the star. For the constant density star above the Buchdahl bound, the inner light ring (iLR) is located at
\begin{align}\label{rilr1}
\rl^{\text{reg}}=\frac{1}{3} \sqrt{ \frac{R_0^3}{GM_0} \frac{4R_0-9GM_0}{R_0-2GM_0}   }\,.
\end{align}
The index `reg' indicates that the metric at this location is regular, in contrast to the situation described below.

The inner light ring is a local \textit{minimum} of the potential $V$, which is guaranteed by topological arguments\,\cite{Cunha:2017qtt, DiFilippo:2024ddg}. This leads to the existence of long lived gravitational modes there, which has been conjectured to lead to instability at the non-linear level:  gravitational waves will focus there, increasing the curvature, which deepens the potential well, leading to further focusing and so on\,\cite{Keir:2014oka}. The conclusion is that stars become dynamically unstable even before approaching the Buchdahl bound. 

What happens if we go beyond the Buchdahl bound? Since we have an explicit solution, we can simply evaluate it in this region of parameters and see what happens. First, notice that the regular inner light ring radius \eqref{rilr1} shrinks to the origin at the Buchdahl limit and becomes imaginary (thus unphysical) beyond that line. However, examining the metric \eqref{lambda}-\eqref{nu} we see that at this point there is a new radius, that is both a light ring ($V$ has a local minimum there) and a curvature singularity, that becomes real beyond this limit:
\begin{align}\label{rilr2}
    \rl^{\text{sing}}=R_0\sqrt{9-\frac{4R_0}{GM_0}}\,.
\end{align}

It is not clear whether this means stability or instability in this case, since the manifold is singular anyways. Anyhow, as we consider solutions that are closer to the black hole limit, the radius of this singular light ring grows outwards until it coincides with the surface ($\rl^{\text{sing}}=R_0$) precisely when $R_0=2GM_0$. 

\begin{figure}[h]
      \centering      \includegraphics[width=.7\columnwidth]{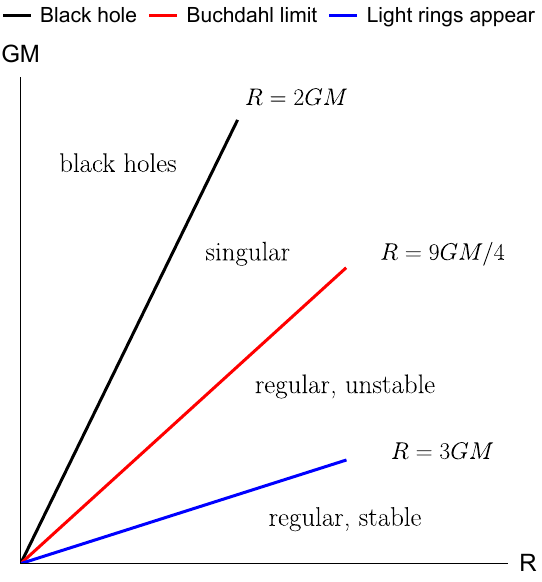}
      \caption{ Space of solutions to the classical problem. Regular means no singularities; unstable means having inner light rings. Angles are not to scale.}
      \label{fig:R_M}
\end{figure}

This limiting solution is an example of a \textit{gravastar}\,\cite{Mazur:2001fv}. As is well known and readily seen from \eqref{p0}, the interior solution becomes a patch of de Sitter with $p=-\rho$ both being constants. This internal dS$_4$ cannot be smoothly connected to an external vacuum Schwarzschild metric since it has a curvature singularity at the surface. In this sense, gravastars are very different from black holes, for the metric is totally smooth at the event horizon. Fig. \ref{fig:R_M} summarizes this section and illustrates the classical picture we have described.  

Having revisited the classical picture, we next define the problem when matter is upgraded to QFT in curved space. We find the exact solution of the fully backreacted system for a perfect fluid star and examine its properties. 

\section{Quantum fields in curved space}

From now on we will focus on semiclassical gravity, where the matter fields are quantized but the metric remains classical. Quantum fields coupled to gauge or gravitational fields exhibit very unique phenomena. In this work we focus on the role of anomalies, in particular the trace, conformal or Weyl anomaly induced by the curvature. Before going into the details of this, let's first consider a more familiar example.  

Consider a free spinor in four dimensional flat space coupled to an electromagnetic field satisfying $\left( i \gamma^\mu D_\mu - m \right)\psi=0$. At the classical level $\partial_\mu J^\mu_5 = 2im \bar\psi \gamma_5\psi$, showing that if the spinor is massless then there is chiral symmetry. However, once the Dirac field is quantized there is an additional contribution from the chiral anomaly \cite{Adler:1969gk,Bell:1969ts}
\begin{align}
   \partial_\mu J^\mu_5  = 2im   \bar\psi \gamma_5 \psi  - \frac{ie^2}{32\pi^2} \epsilon^{\mu\nu\rho\sigma}F_{\mu\nu} F_{\rho\sigma}\,.
\end{align}

This emphasizes that the anomaly is a contribution that comes \textit{on top of} the ones already breaking the symmetry explicitly at the classical level. 

Now let us return the system we are interested in. If the fluid describes a field with conformal symmetry (for example free massless particles) the trace of its stress tensor $(\Tf)^{\mu}_\mu=-\rho+3p$ would vanish. But if one quantizes fields in curved space, the conformal, Weyl or trace anomaly states instead that\,\cite{Deser:1976yx,Duff:2020dqb,Deser:1993yx}
\begin{align}\label{Tmumuq}
    T^\mu_{\ \mu} = -\rho +3p+\frac{1}{(4\pi)^2} \left( c W^2 - a E \right)\,.
\end{align}

Here $W^2=W_{\mu\nu\rho\sigma} W^{\mu\nu\rho\sigma}$ and $E=R^{\mu\nu\rho\sigma} R_{\mu\nu\rho\sigma} - 4R^{\mu\nu} R^{\mu\nu}+R^2$ is the Euler density, which satisfies $\int d^4x \sqrt{-g}E = (2\pi)^2 \chi$, giving the Euler charactersitic when integrated over the manifold. Notice we omitted the term $\Box R$, as its coefficient is scheme-dependent and not universal (more on this below). 

The central charge $c$ and the coefficient $a$ (where we have absorbed $\hbar$) depend on the spin of the fields and are well understood for both the free and interacting cases\,\cite{Duff:2020dqb, Hofman:2008ar}. 
Then just as explained for the chiral anomaly, the conformal or trace anomaly is a contribution that comes on top of those that break conformal symmetry at the classical level. 
This means that the total stress tensor has two components,
\begin{align}\label{Tmunutot}
    T_{\mu\nu}=T_{\mu\nu}{}^{\text{\scriptsize fluid}}+ T_{\mu\nu}{}^{\text{\scriptsize quan}}\,,
\end{align}
where $\Tq$ is defined as those terms generating the anomaly in \eqref{Tmumuq}. But determining the individual components $\Tq_{\mu\nu}$, which encode the effects of QFT coupled to the gravitational field, is where the difficulty lies. All we know in general is its trace. These contributions are notoriously difficult to calculate because they involve renormalization in curved space, and there is no known closed formula for a general metric.

And yet there is one case where this contribution is explicitly known: for conformally coupled quantum fields propagating on a conformally flat (Weyl flat) background. In four dimensions, the result is\,\cite{osti_5430248} 
\begin{align}\label{Tquan}
&T_{\mu\nu}{}^{\text{\scriptsize quant}}= \\
& \frac{-a}{(4\pi)^2} \left[ \left( \frac{ \mathcal{R} ^2}{2}-\mathcal{R}_{\alpha\beta} \mathcal{R}^{\alpha\beta} \right)g_{\mu\nu} + 2 \mathcal{R}_{\mu}^{\ \, \lambda}\mathcal{R}_{\nu\lambda} -\frac{4}{3}\mathcal{R} \mathcal{R}_{\mu\nu}  \right]\,, \nonumber
\end{align}
where $\mathcal{R}_{\alpha\beta}$ is the Ricci tensor, and $a$ is the coefficient appearing in \eqref{Tmumuq}. 

The result \eqref{Tquan} is obtained by functionally integrating the trace anomaly with respect to the conformal factor\,\cite{osti_5430248}. The integration `constant' is then fixed by the requirement that the anomaly vanishes when the spacetime is flat, i.e. $ \langle \Omega | \Tq_{\mu\nu} | \Omega \rangle = 0$ when $g_{\mu\nu}=\eta_{\mu\nu}$. Since this is a local computation, it applies to any conformally flat subset, such as the interior of a star. 

We are finally ready to specify the problem we will consider. We look for solutions to the fully backreacted semiclassical set of Einstein's equations given by:
\begin{align}\label{EEs}
G_{\mu\nu} &= 8\pi G  \left( T_{\mu\nu}{}^{\text{\scriptsize fluid}}+ T_{\mu\nu}{}^{\text{\scriptsize quant}} \right)\\
W_{\mu\nu\rho\sigma}&=0\label{Weyleq}
\end{align}
where $W$ is the Weyl tensor, together with \eqref{metric}, \eqref{Tmunuclass} and \eqref{Tquan}. We emphasize that this system is only self-consistent under the assumption that the resulting metric is Weyl flat, because only then is $\Tq$ given by \eqref{Tquan}. This is the reason why this trick fails in the exterior: the vacuum Schwarzschild solution is not Weyl flat. Furthermore, notice that although \eqref{Tquan} has been evaluated before on top of the fixed Schwarzschild interior star background\cite{PhysRevD.37.2142,Reyes:2023fde}, to the best of our knowledge this is the first time it is used as a source to find the self-consistent solution.    

Since the renormalized quantum stress tensor can be derived from a diffeomorphism invariant effective action for a generic spacetime, it is identically conserved: $\nabla^{\mu} \Tq_{\mu\nu} =0$ (just like the Einstein tensor). Therefore the fluid conservation equation $\nabla^{\mu} T^{\text{fluid}}_{\mu\nu} =0$ leads to the Bianchi identity as usual.

Let us clarify what this system of equations is \textit{not}. It does not correspond to taking the Einstein equations and merely `changing the equation of state', as it is so often done in the literature. Neither is it a higher order derivative theory: the new terms arising from \eqref{Tquan}, albeit non-linear, contain only second derivatives of the metric (this is the other reason to eliminate the $\Box R$ terms in the trace). Rather, the system presented here is a \textit{different set of equations} describing how matter interacts with the geometry, but leaves no room for arbitrary degrees of freedom with respect to General Relativity or QFT.

\section{Semiclassical solution}

We proceed just as in the classical case. Subtracting the radial and angular components of \eqref{EEs} and imposing the Weyl flatness \eqref{noweyl} now gives  
\begin{align}\label{nopressure}
\left( 1-e^{2\lambda}+r\lambda' \right) \left( \pi r e^{2\lambda} +2aG \nu' \right)=0\,,
\end{align}
which together with \eqref{noweyl} forms a closed set of two equations for the two unknowns $\nu,\lambda$.  

This factorization is rather remarkable. In contrast to the classical case, there are \textit{two} families of solutions instead of one. The first branch is identical to the classical equation \eqref{eq000}. And since the Weyl condition \eqref{noweyl} is purely geometric (contains no $a$), it implies that the metric for this branch of exact solutions to the exact equations \eqref{EEs}-\eqref{Weyleq} is identical to the classical one! This is the solution we shall focus on below. The other equation contained in \eqref{nopressure} is briefly commented on in the Discussion. 

As we have just described, the QFT backreacted system of equation admits a solution for the metric that is identical to its classical version: the Schwarzschild interior star \eqref{lambda}-\eqref{nu}. However, even though the metric looks the same as in the classical theory, the physical system is not identical: the semiclassical Einstein equations yield a density and pressure for the fluid different from their classical values:
\begin{align}\label{rhoa}
\rho (r)&=\rho_0  - \frac{8a}{3}G^2\rho_0^2 \\ 
p (r)&=p_0(r) \, + \nonumber \\
&\frac{8a}{3}G^2\rho_0^2  \frac{ \sqrt{R_0^3-2GM_0R_0^2} + \sqrt{R_0^3-2GM_0r^2}  }{\sqrt{R_0^3-2GM_0r^2}-3\sqrt{R_0^3-2GM_0R_0^2}   }\label{pa}
\end{align}
where $\rho_0,p_0$ are given by \eqref{rho0} and \eqref{p0}, respectively.

This is the exact solution to the semiclassical equations for the interior of a star. It is non-perturbative in $a$, as we never assumed it to be small. The energy density is again constant and the fluid isotropic. And although one could renormalise the density, this doesn't work for the pressure because the correction is not proportional to $p_0(r)$. As we explain below, here $M_{0}$ and $R_{0}$ are merely parameters: they do \textit{not} correspond any more to the actual mass and radius of the sphere, to which we turn next.

\subsection{Radius and Mass.} 

The exact solution derived above describes only the interior of the star. In order to construct the entire spacetime, one would like to glue this to an exterior geometry that is also a solution to the semiclassical (not the classical) equations in the vacuum with $\Tf_{\mu\nu}=0$. Such system of equations is currently unknown, let alone its general solution. One would expect that such exterior backreacted metric would be continuously connected to the classical Schwarzschild vacuum solution, but this latter has non-vanishing Weyl tensor, so the method presented here is not useful.

Now, whatever the exterior solution turns out to be, one of its defining properties is that the fluid pressure vanishes outside the star. And similarly to the classical case we require continuity of the pressure at the surface, so that the physical radius $R$ (and not $R_0$) of the sphere is now defined by
\begin{align}\label{pR=0}
p(R)=0\,.
\end{align}

This is the correct condition because $p(r)$ (and nothing else) is the pressure of matter as measured by a device in a sufficiently small region of spacetime. The QFT corrections coming from $\Tq$ should be understood, as shown above, as modifications to the Einstein equations of the system, which evidently affect the numerical value of $p(r)$ as seen in \eqref{pa}. But their effect can only be detected through their interaction with gravity. 

Equation \eqref{pR=0} has a positive solution only if
\begin{align}\label{lambda1/2}
    2aG^2M_0>\pi R_0^3\,.
\end{align}
If so, the positive root is given by
\begin{align}\label{RR0}
R = R_0 \,Q
\end{align}
where
\begin{align}\label{Q2}
Q:=\frac{  \left[ (\pi R_0^3+2aG^2M_0)^2 - 4\pi a G R_0^4 \right]^{1/2}  }{ |\pi R_0^3-2aG^2M_0| }\,.
\end{align}

Defining the Misner mass $m(r)$ in the usual way $e^{-2\lambda(r)}=1-\frac{2Gm(r)}{r}$, the total mass $M:=m(R)$ of the sphere is given by:
\begin{align}\label{MM0}
M = M_0\, Q^3\,.
\end{align}

From \eqref{RR0}-\eqref{MM0} we see that with the QFT effects included, $R_0$ and $M_0$ are \textit{not} the physical radius and mass any more, but rather simply two real numbers parametrizing the space of solutions. The physical radius and mass are $R$ and $M$ respectively.    

In the classical limit $a\to 0$ we have $Q\to 1$ and thus $M\to M_0,R\to R_0$, showing that the semiclassical solution, although fully non-perturbative in $a$, is smoothly connected to the classical one. The Misner mass coincides with the integral $m(r)=-4\pi\int_0^r dr\ r^2\left( T^{0}{}_{0}{}{}^{\text{\scriptsize fluid}}+T^{0}{}_{0}{}{}^{\text{\scriptsize quan}} \right)$. Having found the mass and radius of the spheres, we turn to the analysis of smoothness and stability.

\section{Space of solutions}

In the review of the classical case, we saw that the space of solutions is governed by three important curves in the radius/mass plane: the appearance of the light rings, the Buchdahl bound, and the Schwarzschild limit (corresponding to an interior de Sitter space). For the self-consistent case we can understand the space of solutions by starting from the classical solution and then using the map $(R_0,M_0)\to (R,M)$ defined by \eqref{RR0}-\eqref{MM0}.

\begin{figure}
      \centering      \includegraphics[width=.9\columnwidth]{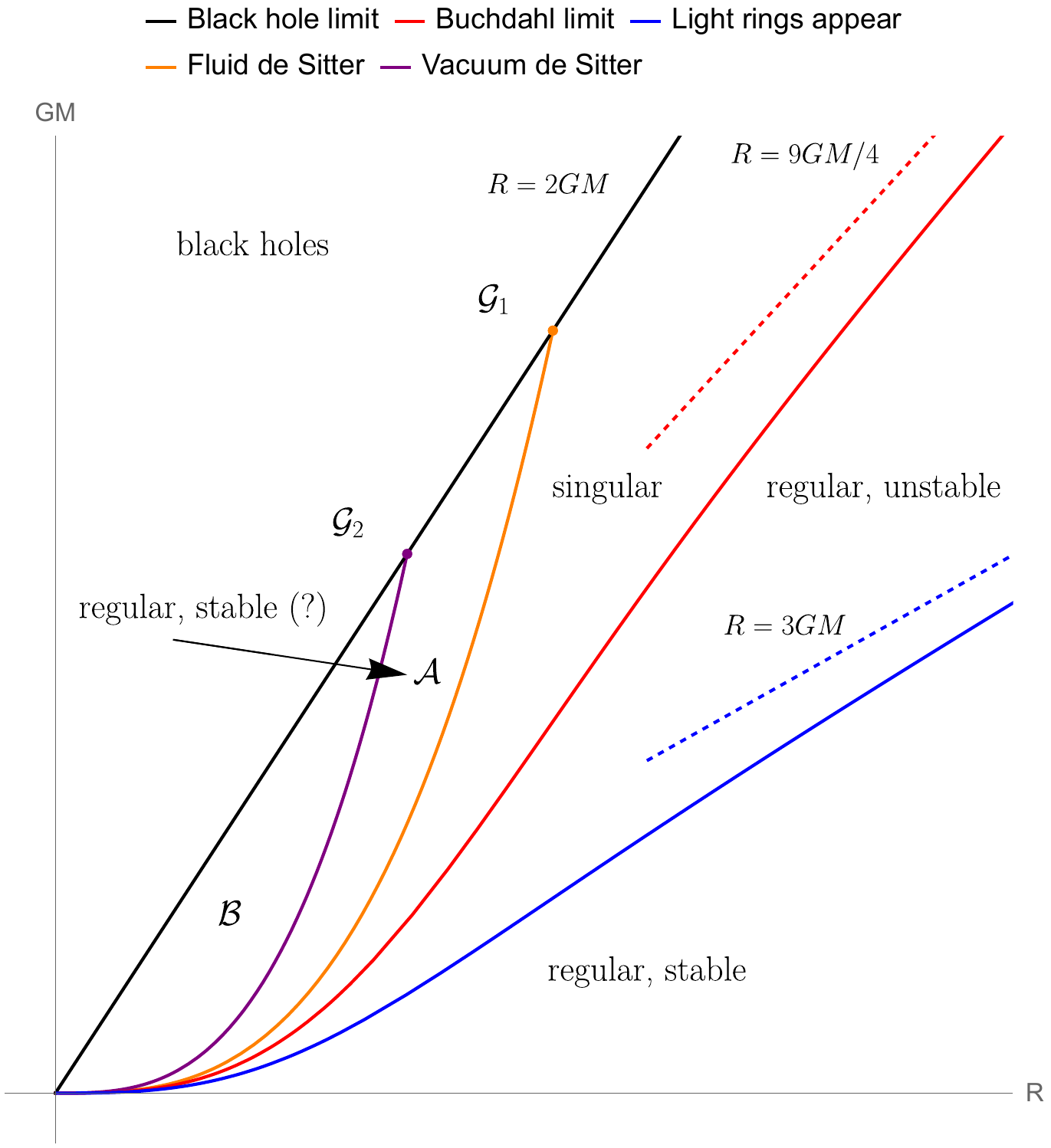}
      \caption{ Space of self-consistent solutions for a  star, the dashed lines indicating their classical counterparts from Fig.\ref{fig:R_M}. $\mathcal{G}_i$ are given by \eqref{G11} and \eqref{G22}. The most novel result is the existence of region $\mathcal{A}$ wherein solutions have no light rings nor singularities inside the star. $\mathcal{B}$ has no star-like solutions. } 
      \label{fig:R_Mq}
\end{figure}

\paragraph{\textbf{Light rings.}}\ Similarly as in the classical case, light rings first appear inside the star when $\rl^{\text{reg}}(R_0,M_0,a)=R(R_0,M_0,a)$, where we have made the dependence explicit to emphasize how this is solved. This defines a curve in the parameter space $(R_0,M_0)$ which via \eqref{RR0}-\eqref{MM0} produces the blue solid curve in the corresponding physical plane $(R,M)$ in Fig. \ref{fig:R_Mq}. 

At very large radius and masses this curve asymptotes to its classical limit (dashed blue), which can be understood by looking at the classical solution. The classical density scales as $\rho_0\sim 1/R_0^2$ along a straight line $M_0\sim R_0$ which vanishes at large $R_0$. For solutions of finite pressure, we have from \eqref{p0} that $p_0(r)\propto \rho_0$ so the pressure become negligible in this limit and thus curvature, and the associated QFT effects, become small in that regime. Here large radius means compared to the scale $\sqrt{a}G\sim \sqrt{a}\ell_{\text P}$ where $\ell_{\text P}$ is the Planck length. We return to this point in the Discussion.    

\paragraph{\textbf{Quantum Buchdahl bound.}}\ One way of characterizing the Buchdahl limit is when the central pressure diverges. By comparing \eqref{p0} and \eqref{pa}, it is clear that the fluid and quantum contributions share the same pole. Therefore the Buchdahl bound is still located at $R_0= \frac{9}{4}GM_0$ in terms of the parameters $R_0,M_0$. But recall that these are \textit{not} any more the physical radius and mass. To find the new Buchdahl bound in terms of the physical values $R,M$, we map this straight line to the physical plane using \eqref{RR0}-\eqref{MM0}. This gives the solid red line in Fig. \ref{fig:R_Mq}. Just like the light ring curve, the quantum bound asymptotes to its classical limit, as it should.  

\paragraph{\textbf{De Sitter solutions.}}\ We saw in the classical case that dS$_4$ space appeared as the interior solution of the star as we approach the Schwarzschild limit (i.e. gravastars). One way of identifying dS space is via the Null Energy Condition (NEC). Choosing the null vector $k^\mu$ along the $(t,r)$ plane, the NEC operator for the fluid is 
\begin{align}\label{}
T^{\text{fluid}}_{\mu\nu} k^\mu k^\nu =\left( \rho+p \right) &= \left( 1- \frac{4aG^2M_0}{\pi R_0^3}  \right) \left( \rho_0+p_0 \right)\,,
\end{align}
where the last term corresponds to the classical NEC. This expression shows that in the QFT case, there are three possible dS$_4$ instead of one.  

The first de Sitter is again the Schwarzschild limit, just as in the classical case. This is particularly simple, as the line $R_0=2GM_0$ is mapped to $R=2GM$. The interior solution is dS$_4$ with $p=-\rho=\frac{3(a-4\pi GM_0^2)}{128 \pi^2 G^4M_0^4}$ corresponding to $\rho_0=-p_0=\frac{3}{32\pi}\frac{1}{G^3M_0^2}$. As we have explained, these solutions should be interpreted as gravastar rather than regular black holes since they satisfy $\rl^{\text{sing}}=R$ so they are singular at the surface. They are depicted again as black solid in Fig. \ref{fig:R_Mq}.

Now we find two dS$_4$ that have no classical analogues. First we have the line $M_0=\frac{\pi R_0^3}{4aG^2}$ in parameter space that maps to another dS$_4$, this time with $\rho=-p=\frac{3}{32 a G^2}$ and corresponds to the solid orange curve in Fig. \ref{fig:R_Mq}. We call these \textit{Fluid de Sitter} solutions, because the density and pressure of the fluid are non-zero. They are singular since again since $\rl^{\text{sing}}=R$. This line ends at the gravastar point 
\begin{align}\label{G11}
   \mathcal{G}_1 : (R,GM) = \sqrt{\frac{aG}{2\pi}} \left( 2,1 \right)\,.
\end{align}

Finally we have the \textit{Vacuum de Sitter} solutions, given by choice 
$(R_0,GM_0) = \sqrt{\frac{aG}{\pi}} \left( 1,1/2 \right)$ corresponding to $\rho_0=\frac{3a}{8G^2}$, a single point in parameter space where the map $(M_0,R_0)\to (M,R)$ becomes ill-defined and \eqref{lambda1/2} is saturated. These solutions have zero matter content since $\rho=0,p=0$ for the fluid (hence the name vacuum), so they are supported only by quantum effects. They are regular: since $\rl^{\text{sing}}>R$ the putative singularity lies outside the surface, where other equations of motion apply, so it is unphysical. Since the fluid pressure vanishes everywhere one can use any radius to define a surface and integrate to get a mass, giving the solid purple curve of Fig. \ref{fig:R_Mq}, ending at the gravastar point  
\begin{align}\label{G22}
    \mathcal{G}_2 : (R,GM) = \sqrt{\frac{aG}{\pi}} \left( 1,\frac{1}{2} \right)\,.
\end{align}

These three dS$_4$ lines define the region $\mathcal{A}$, where solutions have no light rings or singularities inside the star, and the NEC is satisfied. Region $\mathcal{B}$ contains no star-like solutions (the pressure never vanishes) so we discard it.

\section{Discussion}

We have presented an exact non-perturbative solution to the semiclassical equations of the interior of a star which includes the full effects of QFT in curved spacetime, where the metric remains classical and only matter is quantized. We have shown that there is a regime of masses/radii of order $\sim \sqrt{a G}$ where the classical paradigm does not hold, $a$ being the trace anomaly Euler coefficient. There are different possible interpretations for this. 

For the Standard Model $a\sim 1$ so these objects are of Planckian size. In the absence of a satisfactory theory of quantum gravity (which may not even exist), we must either take this at face value or declare the result invalid. On the other hand one can consider this solution in the `academic' regime $a\gg 1$ where no Planckian scales are involved. Either way, the result stands as the exact solution to the well defined mathematical problem of self-consistent semiclassical equations. 

The existence of the region $\mathcal{A}$ bounded by three dS$_4$ lines is one of the main results of this work. By definition, the interior of stars in $\mathcal{A}$  contain no singularities. Nor can they contain any light rings: for those geometries the two candidate radii \eqref{rilr1} and \eqref{rilr2} are either imaginary or lie outside the surface, respectively, so they are unphysical. Solutions in this region are the most incompatible with the standard theory, which claims that configurations sufficiently close to the Schwarzschild limit will either be singular or have interior light rings. 

This leads to a striking conclusion about the exterior geometry of solutions in $\mathcal{A}$. Assuming they are also smooth manifolds, topological theorems guarantee that light rings come in min/max pairs\,\cite{Cunha:2017qtt,DiFilippo:2024ddg}. And since there's no light rings in the interior, we conclude that \textit{the exterior semiclassical solution must have either no light rings or an even number of them}. 

Finally as we noted above, the semiclassical equation \eqref{nopressure} possesses, in addition to a star-like one, a second branch of solutions satisfying $\pi r e^{2\lambda} +2aG \nu'=0$,
which introduced back into \eqref{noweyl} leads to:
\begin{align}\label{}
r\frac{\nu''}{\nu'} + r^2\nu''+2r^2(\nu')^2 + \left( \frac{4aG}{\pi r}-r \right)\nu'+1=0\,.
\end{align}

This branch only exists for $a>0$, so it is a genuinely QFT effect with no classical counterpart. It does not describe an object with a physical surface, and a thorough analysis goes beyond the scope of this work.

We finish by suggesting some open questions and ideas. First, the result\,\eqref{Tquan} used here captures only `local' geometric terms, but the effect of the more general non-local contributions\,\cite{Riegert:1984kt,PhysRevD.21.2185,PhysRevD.15.2088} remains a very important open problem (see also \cite{Hollands:2020qpe,Zilberman:2019buh,Arrechea:2023hmo} regarding horizons).  And connected to this: what are the right semiclassical equations (let alone their solution) for a self-consistent QFT backreacted vacuum exterior? Do these exterior solutions, in particular those associated with configurations in region $\mathcal{A}$ of Fig. \ref{fig:R_Mq}, have light rings or not? And if such novel solutions in $\mathcal{A}$ were to exist in Nature, how would one detect them? Last but not the least, all gravitational anomalies in $2n$ dimensions derive from an Atiya-Singer index theorem in $2n+2$ dimensions\,\cite{1985AnPhy.161..423A, Alvarez-Gaume:1983ihn}. It appears that one could make more progress by taking this as starting point. 

\section{Acknowledgements}

We are grateful to Julio Arrechea, Valentin Boyanov, Carlos Barcel\'o, Jan de Boer, Ben Freivogel, Luis Garay, Gerardo Garc\'ia-Moreno, Bahman Najian, Banafshe Shiralilou and especially Gimmy Tomaselli for insightful discussions.  

\bibliography{Full.bib}

\end{document}